\documentclass[conference, a4paper]{IEEEtran}
\usepackage[T1]{fontenc}
\IEEEoverridecommandlockouts
\makeatletter
\def\ps@headings{
\def\@oddhead{\mbox{}\scriptsize\rightmark \hfil \thepage}%
\def\@evenhead{\scriptsize\thepage \hfil \leftmark\mbox{}}%
\def\@oddfoot{}
\def\@evenfoot{}}
\makeatother
\usepackage{subfigure}
\usepackage{colortbl}
\usepackage{bm}
\usepackage{graphicx}
\usepackage{epstopdf}

\usepackage{graphicx}  % Written by David Carlisle and Sebastian Rahtz
\usepackage{url}       % Written by Donald Arseneau

\usepackage{amsmath}   % From the American Mathematical Society
\usepackage{cite}
\usepackage{extarrows}
\usepackage{amsfonts,amssymb}

\usepackage{stfloats}

\usepackage{cases}
\usepackage{algorithm}
\usepackage{multirow}
\usepackage{algorithmic}
\newcommand{\ls}[1]
    {\dimen0=\fontdimen6\the\font
     \lineskip=#1\dimen0
     \advance\lineskip.5\fontdimen5\the\font
     \advance\lineskip-\dimen0
     \lineskiplimit=.9\lineskip
     \baselineskip=\lineskip
     \advance\baselineskip\dimen0
     \normallineskip\lineskip
     \normallineskiplimit\lineskiplimit
     \normalbaselineskip\baselineskip
     \ignorespaces
    }

\begin{document}

\title{Dumb RIS-Assisted Random Beamforming\\ for Energy Efficiency Enhancement\\ of Wireless Communications}

\author{\IEEEauthorblockN{Yixin Zhang$^{\dagger}$, Wenchi Cheng$^{\dagger}$, and Wei Zhang$^{\ddagger}$}~\\[0.2cm]
\vspace{-10pt}

\IEEEauthorblockA{$^{\dagger}$State Key Laboratory of Integrated
Services Networks, Xidian University, Xi'an, China\\
$^{\ddagger}$School of Electrical Engineering and Telecommunications, The University of New South Wales, Sydney, Australia\\
E-mail: \{\emph{yixinzhang@stu.xidian.edu.cn}, \emph{wccheng@xidian.edu.cn},
\emph{w.zhang@unsw.edu.au}\}}}

\vspace{-20pt}

\maketitle

\begin{abstract}
Energy efficiency (EE) is one of the most important metrics for the beyond fifth generation (B5G) and the future sixth generation (6G) wireless networks. Reconfigurable intelligent surface (RIS) has been widely focused on EE enhancement for wireless networks because it is low power consuming, programmable, and easy to be deployed. However, RIS is generally passive and thus difficult to obtain corresponding full channel state information (CSI) via RIS, which severely impacts the EE enhancement of RIS-assisted wireless communications. To solve this problem, in this paper we propose the new single-active-antenna combined RIS transmitter structure, which can replace traditional multiple antennas to reduce hardware cost and power consumption. Based on the single-active-antenna combined RIS structure, we develop the Dumb RIS-Assisted Random Beamforming (Darb)-based Joint RIS-Elements and Transmit-power optimizAtion (Jeta) scheme, where dumb RIS randomly changes its phase shift according to isotropic distribution only depending on the CSI feedback from users to RIS-assisted transmitter. Then, we jointly design the number of RIS elements and optimize the transmit power to maximize the EE of RIS-assisted wireless communications. Simulation results show that compared with the traditional multi-antenna system, our developed Darb-based-Jeta scheme can significantly increase the EE without the full CSI.
\end{abstract}

\vspace{10pt}


%\IEEEpeerreviewmaketitle

\section{Introduction}
\IEEEPARstart{D}{etermined} by the demand for higher rates, network capacity is required to continue to increase by 1000 times in the future wireless networks~\cite{6G}. In the beyond fifth generation (B5G) and the sixth generation (6G) communication systems, various promising technologies such as ultra-massive multiple-input multiple-output (UM-MIMO) and terahertz (THz) communications are expected to achieve a high access rate and network capacity. However, they require a large number of radio frequency (RF) chains and result in high hardware cost and complexity~\cite{RFC}. As a result, energy consumption remains a difficult problem and cost-effective solutions are still very important for future wireless networks.

Recently, reconfigurable intelligent surface (RIS) has been proposed as a potential technology to provide a new possibility for energy-efficient wireless networks. RIS is a plane composed of a large number of low-cost passive reflecting elements. Each element can independently reflect the incident signal by controlling its amplitude and phase, thereby cooperatively realizing reflecting beamforming. Since the RIS is passive, it does not depend on RF chains and its energy consumption is very low, enabling low-cost, low power consuming and ultra-dense deployment~\cite{WuMag}.

Relevant work has been carried out to study the energy efficiency (EE) optimization problem in the RIS-assisted communication systems~\cite{EE-Huang,EE-CL,EE-SE}. A practical RIS power consumption model was developed with jointly optimizing the RIS phase shift and the downlink transmit power to maximize the EE under quality of service (QoS) constraints in the RIS-aided downlink multi-user multiple-input single-output (MISO) system~\cite{EE-Huang}. Taking the limited backhaul capacity constraints into account,  the authors proposed a joint design of transmit beamforming and reflecting coefficients at RISs to maximize the EE of cell-free networks~\cite{EE-CL}. The authors studied the tradeoff between EE and spectrum efficiency (SE) in RIS-aided multi-user MIMO uplink systems~\cite{EE-SE}.

Even though various investigations on the benefits of using RIS for green communications have been carried out, they are almost to design passive reflecting beamforming under the assumption that the channel state information (CSI) is perfectly known at the RIS side. However, since the RIS is passive and with relatively low signal processing capability, the perfect CSI is very challenging to be obtained at the RIS side. In addition, because a RIS controls a significant number of elements, the number of channels to be estimated is large~\cite{CSI-JSAC}. All these factors have brought the challenge to the channel estimation in the RIS-assisted system, which severely impacts the EE enhancement of RIS-assisted wireless communications.

In order to solve this problem, in this paper we place a RIS near the RF signal generator to form the new single-active-antenna combined RIS transmitter structure, which can replace multiple antennas to reduce hardware cost and power consumption. Based on the single-active-antenna combined RIS structure, we develop the dumb RIS-assisted random beamforming (Darb) scheme with a threshold feedback strategy, where dumb RIS randomly changes its phase shift according to isotropic distribution. As a result, the phase shift of the RIS is randomly generated by the RIS controller so that optimal phase shift is not required to design. The RIS-assisted system only depends on the overall channel-related feedback from users to the RIS-assisted transmitter. Thus, the problem of obtaining full CSI via RIS is avoided. Then, we jointly design the number of RIS elements and the transmit power to optimize the EE of RIS-assisted wireless communication system. The optimization problem is a mixed-integer non-convex problem that is difficult to solve. As such, we propose an alternating optimization algorithm for solving it, where the number of RIS elements and the transmit power are alternately optimized in each iteration until the result converges. Simulation results show that our developed Darb-based joint RIS-elements and transmit-power optimization (Jeta) scheme can significantly increase the EE without the need to know the full CSI.

The rest of this paper is organized as follows. Section~\ref{sec:sys} introduces the new RIS-assisted transmitter structure and the signal model. Section~\ref{sec:III} presents the power consumption and the energy efficiency model. Section~\ref{sec:O} presents the Darb-based-Jeta scheme. Section~\ref{sec:simu} provides the numerical results. Finally, we conclude this paper in Section~\ref{sec:conc}.

\section{System Model}\label{sec:sys}

\addtolength{\topmargin}{0.27in}

%原来是{S.eps}
\subsection{RIS-Assisted Transmitter Structure}
\begin{figure}[htbp]
\centering\includegraphics[width=3.5in]{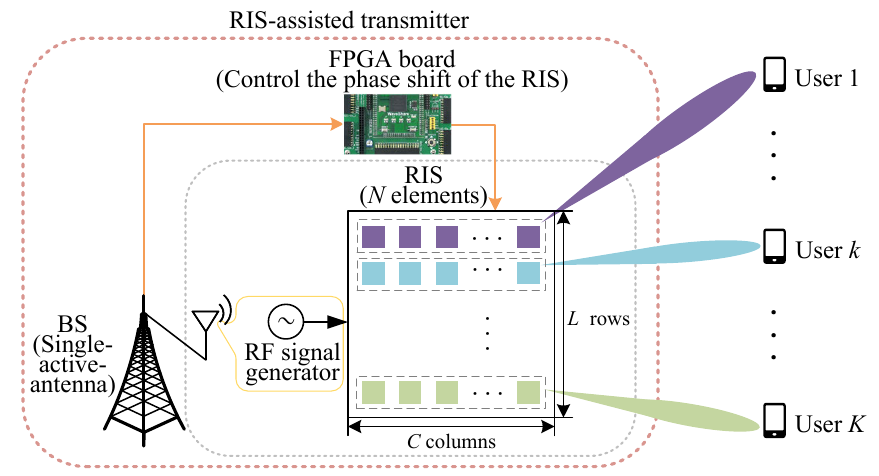}
\caption{The RIS-assisted wireless communications system.}\label{fig:1}
\end{figure}
Figure~\ref{fig:1} shows the new single-active-antenna RIS-assisted transmitter structure-based wireless communications system, where a RIS is placed very close to an antenna. The RF signal generator of the antenna transmits an unmodulated carrier signal to the RIS. Through the adjustment of phases on the RIS, user information can be  transmitted. The distance between the RIS and the RF signal generator is very small so that the transmission is not affected by fading~\cite{8}. The RIS consists of $L$ rows and $C$ columns so that $N=LC$. The RIS element in the $l$-th row and the $c$-th column is denoted by $E_{lc}$ and controlled by a field programmable gate array (FPGA) board to change its phase shift.

In this paper, to provide multiple users with multiple data streams, we divide the RIS into different regions by row to realize the function of multiple antennas. That is, the $C$ elements in the $l$-th row of the RIS are assigned to a user $k, k=1,2,\cdots, K$, i.e., the phases of RIS elements $E_{l1},E_{l2},\cdots,E_{lC}$ are used to control the transmit signal for user $k$. Compared with traditional BS with multiple transmit antennas, the RIS-assisted transmitter only needs one active antenna. Therefore, the RIS-assisted transmitter can reduce hardware cost and power consumption to achieve high energy efficiency.

\subsection{Signal Model}
We consider a RIS-assisted multi-user MIMO broadcast channel, where a single-active-antenna base station (BS) and a RIS with $N$ passive reflecting elements are used together to transmit common information to $K$ single-antenna users randomly located on the ground. In this paper, we assume the downlink communication undergoes block fading and the channel stays constant during a time-slot $t$ of length $T$ corresponding to the coherence interval. In order to serve multiple users in the same time-slot, a precoding matrix $\bm W\in\mathbb{C}^{L\times C}$ is applied at the RIS-assisted transmitter. The RIS element of each row can be regarded as a beam, which performs the same function as an antenna. $\bm W=[\bm w_1,\bm w_2,\cdots,\bm w_L]^T$, where $\bm w_l\in\mathbb{C}^{1\times C}, l=1,2,\cdots,L,$ denotes the phase shift of RIS elements on the $l$-th row.

Since the signal reflected from the RIS has experienced high path loss, we only consider the first reflected signal and ignore the power of the signal reflected twice or more from the RIS \cite{Wu2}. Then, the signal received at user $k$, with $k=1,2,\cdots,K$, is written as
\begin{equation}
\begin{aligned}
y_k&=\sum_{l=1}^{L}\sqrt{p_l}\bm h_k^T \bm w_l s_l+n_k\\
&=\sqrt{p_i}\bm h_{k}^T \bm w_i s_i+\sum_{l=1,l\neq i}^{L}\sqrt{p_l}\bm h_{k}^T \bm w_l s_l+n_k,
\end{aligned}
\end{equation}
where $p_l$ is the transmit power for beam $l$, $\bm h_k \in\mathbb{C}^{C\times 1} $ is the complex Gaussian channel vector between the RIS-assisted transmitter and the $k$-th user, $s_l$ is the signal corresponding to beam $l$ and $n_k\sim \mathcal{CN}(0,\sigma^2)$ is the additive white Gaussian noise (AWGN) at the $k$-th user. All users are assumed to experience independent and identically distributed (i.i.d.) Rayleigh fading $\bm h_k\sim \mathcal{CN}(0,\beta\bm I_C)$, where $\beta$ denotes path loss. We assume that the total transmit power $P_T$ is equally allocated to each beam, i.e., $p_l=\frac{P_T}{L}, l=1,2,\cdots,L$. We assume that the $k$-th user knows $\bm h_k^T \bm w_l$ for $l=1,2,\cdots,L$. There is an error-free and delay-free feedback channel that can pass part of the CSI to the RIS-transmitter. Hence, the corresponding received signal-to-interference and noise ratio (SINR) for the $k$-th user with the $i$-th beam is given as
\begin{equation}
\gamma_{k,i}=\frac{|\bm h_{k}^T\bm w_i|^2}{\sum_{l=1,l\neq i}^{L}|\bm h_{k}^T\bm w_l|^2+\frac{L\sigma^2}{P_T}}.
\end{equation}

A max-SINR rule is used to coordinate the scheduling process. That is, each user feedbacks  its highest SINR among all beams and the corresponding beam index to the RIS-assisted transmitter. Then, the RIS-assisted transmitter selects the user with the highest SINR for each beam to maximize the sum rate. Therefore, the sum rate of the RIS-assisted system, denoted by $R$, can be expressed as follows:
\begin{equation}
R=L\mathbb{E} \Big\{ \log_2(1+\mathop{\max}\limits_{1\leq k \leq K} \gamma_{k,i})    \Big\}.
\end{equation}

\section{Problem Formulation} \label{sec:III}
In this section, we analyze the total power consumption in both RIS-assisted and multi-antenna systems and compare their energy efficiencies. For the RIS-assisted system, we formulate the energy efficiency optimization problem with the joint design of the number of RIS elements and the transmit power.
\subsection{Total Power Consumption and Energy Efficiency Model}
The power consumption of RIS, denoted by $P_{RIS}$, can be given as follows:
\begin{equation}
P_{RIS}={P_{FPGA}+NP_{PIN}},
\end{equation}
where $P_{FPGA}$ and $P_{PIN}$ are the power consumption for the FPGA board and the PIN diode on RIS, respectively. Since the RIS-assisted transmitter only has one active-antenna, the circuit power consumption of the RIS-assisted system can be given by
\begin{equation}
P_{CR}=P_A+LP_U,
\end{equation}
where $P_A$ is the circuit power consumed by the single-active-antenna, consisting of digital-to-analog converter (DAC), mixer and filters of transmitter~\cite{CuiEE}. While $P_U$ is the circuit power consumed at the user side, which contains low noise amplifier, mixer, filters of the receiver and analog-to-digital converter (ADC). Thus, the total power consumption of the RIS-assisted system can be derived as follows:
\begin{equation}
P_{RA}=\frac{P_T}{\eta_T}+P_{RIS}+P_{CR}+P_{SR}+\sum_{k=1}^K P_{U,k},
\end{equation}
where $\eta_{T}$ is the power conversion efficiency of transmit power amplifier, $P_{SR}$ and $P_{U,k}$ are the hardware static power of the RIS-assisted transmitter and the $k$-th user equipment, respectively.

On the other hand, the total power consumption of the multi-antenna transmitter system, denoted by $P_{MA}$, is given by
\begin{equation}
P_{MA}=\frac{P_T}{\eta_T}+P_{CA}+P_{SA}+\sum_{k=1}^K P_{U,k},
\end{equation}
where $P_{CA}=MP_A+MP_U$ is the circuit power consumption of the multi-antenna system, $M$ is the number of antennas, and $P_{SA}$ is the hardware static power of the multi-antenna transmitter. To fairly compare with RIS-assisted transmitter, we set $M=L$.

Then, we have the energy efficiencies of RIS-assisted and multi-antenna system, denoted by $EE_{RA}$ and $EE_{MA}$, respectively, as follows:
\begin{equation}
\left\{
\begin{aligned}
EE_{RIS}=\frac{R}{P_{RA}}, & \rm{\ for\  RIS-assisted\  system;}\\
EE_{MA}=\frac{R}{P_{MA}}, & \rm{\ for\  multi-antenna\  system.}
\end{aligned}
\right.
\end{equation}
\subsection{Energy Efficiency Optimization Problem Formulation}
In order to increase the energy efficiency of the RIS-assisted wireless communication system, we propose the energy efficiency optimization problem by jointly optimizing the number of RIS elements $N$ and the transmit power $P_T$. We set the maximum number of RIS elements as $N_{\max}$ and the maximum transmit power as $P_{\max}$. The joint optimization problem, denoted by  $\textbf{\textit{P}1}$, is formulated as follows:

\begin{align}
\textbf{\textit{P}1:\ } \max_{\left(N, P_T\right) }\ \ &\frac{L\mathbb{E} \Big\{ \log_2(1+\mathop{\max}\limits_{1\leq k \leq K} \gamma_{k,i})    \Big\}}{\frac{P_T}{\eta_T}+P_{RIS}+P_{CR}+P_{SR}+\sum_{k=1}^K P_{U,k}}\\
\  \mathrm{s.t.}\ \  &1).\ N\in\{1,2,\cdots,N_{\max}\};\label{integer}\\
\ \  &2).\ 0<P_T\leq P_{\max}.
\end{align}

\section{Darb-based Energy Efficiency Optimization}\label{sec:O}
The full CSI of the RIS-assisted system is difficult to be obtained, which impacts the energy efficiency enhancement. In order to achieve high energy efficiency without full CSI, we propose the Darb-based-Jeta scheme.

\subsection{Random Beamforming Using Dumb RIS}
In order to optimize the energy efficiency of the RIS-assisted system, the phase shift needs to be optimally changed. However, the optimal phase shift control of RIS requires perfect CSI of all links between the RIS-assisted transmitter and the users, which is very difficult to be obtained because RIS is passive. Therefore, it is necessary to perform channel estimation and corresponding feedback mechanism at the RIS-assisted transmitter and the user side. Motivated by opportunistic beamforming~\cite{OBF}, we propose the dumb RIS-assisted random beamforming (Darb) scheme that constructs random beams without full CSI at the RIS-assisted transmitter side. We use the dumb RIS, where the word dumb means not to adjust any phase shift on the RIS but to let it change randomly. That is to say, the phase of RIS does not require any control, so there is no need to optimize it and no need to know the full CSI.

According to the orthogonal random beamforming strategy for multi-user transmission~\cite{ORB}, we generate a random unitary matrix $\bm \Phi = [\bm\phi_1, \bm\phi_2,\cdots, \bm\phi_{L}]\in\mathbb{C}^{L\times L}$ on RIS in each time-slot to transmit $L$ signals simultaneously, where $\bm \phi_l = [e^{j\theta_{1l}}, e^{j\theta_{2l}},\cdots,e^{j\theta_{Ll}}]^T \in \mathbb{C}^{L\times 1}, l = 1,\cdots, L,$ is orthonormal vector generated from an isotropic distribution randomly \cite{IS}. Then, we set the number of rows and columns in RIS to be the same value $L$, i.e., $N=L^2$. Hence, the corresponding received SINR for the $k$-th user with the $i$-th beam can be rewritten as follows:
\begin{equation}
\gamma_{k,i}=\frac{|\bm h_{k}^T\bm \phi_i|^2}{\sum_{l=1,l\neq i}^{L}|\bm h_{k}^T\bm \phi_l|^2+L/\rho}=\frac{z}{y+\frac{L\sigma^2}{P_T}}.
\end{equation}
Since the channels of all users are assumed to be i.i.d. Rayleigh channels and $\bm \Phi$ is a unitary matrix, the two variables $z$ and $y$ are independent chi-square distributed with $z\sim \mathcal{X}^2(2)$, $y\sim \mathcal{X}^2(2L-2)$, respectively. Then, the cumulative density function of the SINR, $\forall k,i$, can be expressed as \cite{ORB}
\begin{equation}\label{CDF}
F(\gamma) = 1-\frac{e^{- \frac{ L \sigma^2\gamma}{P_T}}}{{(1 + \gamma)}^{L-1}}, \ \gamma\geq 0.
\end{equation}
According to the max-SINR rule, the selected user $k_i^*$ for beam $i$ is given by
\begin{equation}
k_i^* = \mathop{\arg\ \max}_{1\leq k \leq K}\ \gamma_{k,i}, \ \ i = 1,2,\cdots,L.
\end{equation}
Then, the probability density function of the selected user $k_i^*$'s SINR is given by  \cite{ORB} 
\begin{equation}
\begin{aligned}
f_{k^*}(\gamma) = K\frac{e^{- \frac{L\sigma^2\gamma }{P_T}}}{(1+\gamma)^L}&{\Big [\frac{L\sigma^2}{P_T}(1 + \gamma)+ L-1}\Big ]
\\& \times \Big[1-\frac{e^{- \frac{L\sigma^2\gamma }{P_T}}}{{(1 + \gamma)}^{L-1}}\Big]^{K-1}.
\end{aligned}
\end{equation}
The RIS-assisted transmitter forms $L$ groups according to the beam index and the user who has the highest SINR in each group is chosen to be transmitted. Finally, the sum rate of the Darb scheme, denoted by $R_{Darb}$, can be expressed as follows:
\begin{equation}
R_{Darb} = L\int_0^\infty \log_2(1+\gamma)f_{k^*}(\gamma) d\gamma.
\end{equation}
As the number of users grows to infinity, the sum rate of the Darb scheme can be reached as follows:
\begin{equation}\label{sumrate}
R_{Darb}\mathop{=}^{K\rightarrow \infty}L\log(\beta\log K)+L\log\frac{P_T}{L\sigma^2}.
\end{equation}
This means that when the number of users is large enough, the sum rate of the Darb scheme is the same as the sum rate be achieved by dirty paper coding~\cite{DPC}. The required feedbacks are only the highest SINRs of each user and the corresponding beam index, which means RIS-assisted transmitter does not need full CSI. The passive channel estimation via RIS is avoided, which reduces the overhead of CSI acquisition and simplifies the RIS design.

When the number of users $K$ in the system is large, the feedback overhead that linearly increases with $K$ is also large, which is given by
\begin{equation}
FO=\underbrace{KQ}_{\text{the highest SINR}}+\underbrace{K\log_2{L}}_{\text{corresponding beam index}},
\end{equation}
where $Q$ is the quantization bits of the highest SINR. In order to reduce large feedback overhead, we set a threshold $\alpha$ on each row. Each user calculates its SINR on each beam and obtains the highest value from it. Then, it compares the highest SINR with the given threshold $\alpha$. If the highest SINR of the user is larger than the given threshold $\alpha$, the user feedbacks the SINR and the corresponding beam index. Otherwise, the user does not feedback information. Thus, the sum rate of Darb scheme with the threshold feedback strategy can be expressed as follows:
\begin{equation}
R_{TFS}=\Big[1-F^K(\alpha)\Big]R_{Darb}.
\end{equation}
When we appropriately choose the threshold $\alpha$, it has
\begin{equation}
\lim_{K\rightarrow\infty}\frac{R_{TFS}}{R_{Darb}}=1.
\end{equation}
Thus, the capacity loss caused by the threshold feedback strategy is small. The feedback overhead of the Darb scheme with the threshold feedback strategy is given by
\begin{equation}
FO_{TFS}=\Big[1-F(\alpha) \Big]FO.
\end{equation}

\subsection{Darb-based Joint RIS-elements and Transmit-power Optimization Scheme}
Based on the Darb scheme, we can rewrite the energy efficiency of the RIS-assisted system, denoted by $EE_{Darb}$, as follows:
\begin{equation}
EE_{Darb}=\frac{L\log(\beta\log K)+L\log\frac{P_T}{L\sigma^2}}{\frac{P_T}{\eta_T}+L^2P_{PIN}+LP_{U}+P_1},
\end{equation}
where $P_1=P_{FPGA}+P_{A}+P_{SR}+\sum_{k=1}^K P_{U,k}$. By setting the maximum number of RIS rows as $L_{\max}$, we can convert problem $\textbf{\textit{P}1}$ to problem $\textbf{\textit{P}2}$ as follows:
\begin{align}
\textbf{\textit{P}2:\ } \max_{\left(L, P_T\right) }\ \ &EE_{Darb}\\
\  \mathrm{s.t.}\ \  &1).\ L\in\{1,2,\cdots,L_{\max}\};\label{integer}\\
\ \  &2).\ 0<P_T\leq P_{\max}.
\end{align}
However, note that problem $\textbf{\textit{P}2}$ is a mixed-integer non-convex optimization problem since Eq.~(\ref{integer}) involves integer constraint while the two optimization variables $L$ and $P_T$ are coupled with each other. To decouple $L$ and $P_T$, we apply the alternating optimization (AO) algorithm to divide the problem into two sub-problems and optimize them separately. Specifically, one optimizes $L$ with fixed $P_T$ and the other optimizes $P_T$ with fixed $L$.%

First, we fix the transmit power $P_T$ and optimize the number of RIS elements $L^2$. In order to make problem $\textbf{\textit{P}2}$ be conveniently solved, we relax the discrete variables in Eq.~(\ref{integer}) into continuous variables. Then, the sub-problem $\textbf{\textit{P}3}$ is formed as follows:
\begin{eqnarray}
\hspace{-3.9em}\textbf{\textit{P}3:} \ \ \ \ \   \max_{\left(L\right) }&  &EE_{Darb} \label{L}\\
\  \mathrm{s.t.}& & 1\leq L\leq L_{\max}. \label{eq:P1_c_1}
\end{eqnarray}
Define $f(l)=\displaystyle{\frac{l\log(\beta\log K)+l\log\frac{P_T}{l\sigma^2}}{l^2P_{PIN}+lP_U+a}}$, where $\displaystyle{a=\frac{P_T}{\eta_T}+P_{FPGA}+P_A+P_{SR}+\sum_{k=1}^K P_{U,k}}$ is a constant, for all $l_1,l_2 \in L$ and $\nabla f(l_1)(l_2-l_1)\geq 0$, satisfying $f(l_2) \geq f(l_1)$. Therefore, Eq.~(\ref{L}) is a strictly pseudo-concave function. The strictly pseudo-concave function is monotonically increasing, or it has a unique stationary point which is also its global maximum point~\cite{1975Pseudo}. Thus, we can optimize $L$ by finding its unique stationary point.

Then, we fix the number of RIS elements $L^2$ and optimize the transmit power $P_T$. The sub-problem $\textbf{\textit{P}4}$ is formed as follows:
\begin{eqnarray}
\hspace{-3.9em}\textbf{\textit{P}4:} \ \ \ \ \   \max_{\left(P_T\right) }&  &EE_{Darb}\label{PT}\\
\  \mathrm{s.t.}& & 0 < P_T\leq P_{\max}. \label{eq:P1_c_1}
\end{eqnarray}
Define $f(p)=\displaystyle{\frac{L\log\frac{p}{L\sigma^2}+b}{\frac{p}{\eta_T}+c}}$, where $b=L\log(\beta\log K)$ and $c=P_{CA}+P_{SA}+\sum_{k=1}^K P_{U,k}$ are constants. The numerator of $f(y)$ is strictly concave and the denominator is an affine. The ratio of a strictly concave function to an affine function is also a strictly pseudo-concave function, which is easy to find its global maximum point.

\subsection{Darb-based-Jeta Scheme and Its Convergence}
Based on the results of two sub-problems, the alternating optimization is adopted to solve problem $\textbf{\textit{P}2}$. Specifically, in each iteration, while keeping the other variable fixed, the number of RIS elements $L^2$ and the transmit power $P_T$ are alternately optimized by solving sub-problems $\textbf{\textit{P}3}$ and $\textbf{\textit{P}4}$. Furthermore, the solution obtained in each iteration is used as the input of the next iteration. The specific algorithm is given in Algorithm 1.

\begin{algorithm}[h]
\caption{Alternating optimization for solving (P2).}
\begin{algorithmic}[1]
\STATE Initialize iteration index $t=0$.
\STATE Initialize the the number of RIS rows $L^{(0)}=1$ and the transmit power $P_T^{(0)}=0$, where the label in the upper right corner represents the number of iterations.
\REPEAT
\STATE For the given transmit power $P_T^{(t)}$, update the number of RIS row as $L^{(t+1)}$ by solving problem $\textbf{\textit{P}3}$.
\STATE For the given number of RIS rows $L^{(t+1)}$, update the transmit power as $P_T^{(t+1)}$ by solving problem $\textbf{\textit{P}4}$.
\STATE Update $t\leftarrow t+1$.
\UNTIL{the increment of $EE_{Darb}$ is smaller than the given threshold $\epsilon$.}
\label{code:recentEnd}
\end{algorithmic}
\end{algorithm}

Next, we prove the convergence of Algorithm 1 as follows:
\begin{equation}
\begin{aligned}
EE_{Darb}\left(L^{(t)},P_T^{(t)} \right)&\overset{(a)}\leq EE_{Darb}\left(L^{(t+1)},P_T^{(t)}\right)\\
&\overset{(b)}\leq EE_{Darb}\left(L^{(t+1)},P_T^{(t+1)}\right),
\end{aligned}
\end{equation}
where (a) holds in step 4 of Algorithm 1 and (b) holds in step 5. The objective value of problem (P1) is non-decreasing, which guarantees the convergence of Algorithm 1.

\section{Performance Evaluations}\label{sec:simu}
In this section, we evaluate the performance of our proposed Darb-based-Jeta scheme for RIS-assisted wireless communications. We place the RIS-assisted transmitter at the origin and users are randomly distributed in a square area with a side length of $60$~m. We set the bandwidth $B = 180$~KHz, the noise variance $\sigma^2 = -80$~dBm, the path loss from RIS-assisted transmitter to user $\beta = \frac{10^{-3.53}}{d^{3.76}}$, the quantization bits of the user's highest SINR $Q=4$ and the algorithmic convergence parameter $\epsilon = 0.05$, respectively. The parameters of power consumption are shown in Table \ref{Parameters} \cite{CuiEE}.
\begin{table}[htbp]
 \setlength{\tabcolsep}{5mm}
\caption{\label{Parameters}Simulation Parameters}
\centering
\renewcommand\arraystretch{1.5}
\begin{tabular}{|c|c|c|c|}
	\hline Parameters&Values&Parameters&Values\\[2.8pt]
	\hline $P_{\max}$&13dBW&$\eta_{T}$&80\%\\
	\hline $P_{PIN}$&7dBm&$P_{FPGA}$&27dBm\\
	\hline $P_{A}$&20dBm&$P_{U}$&20dBm\\
	\hline $P_{SR}$&30dBm&$P_{SA}$&33dBm\\
	\hline $P_{U,k}$&10dBm&$L_{\max}$&20\\
	\hline
\end{tabular}
\end{table}

\begin{figure}[htbp]
\centering\includegraphics[width=3.4in]{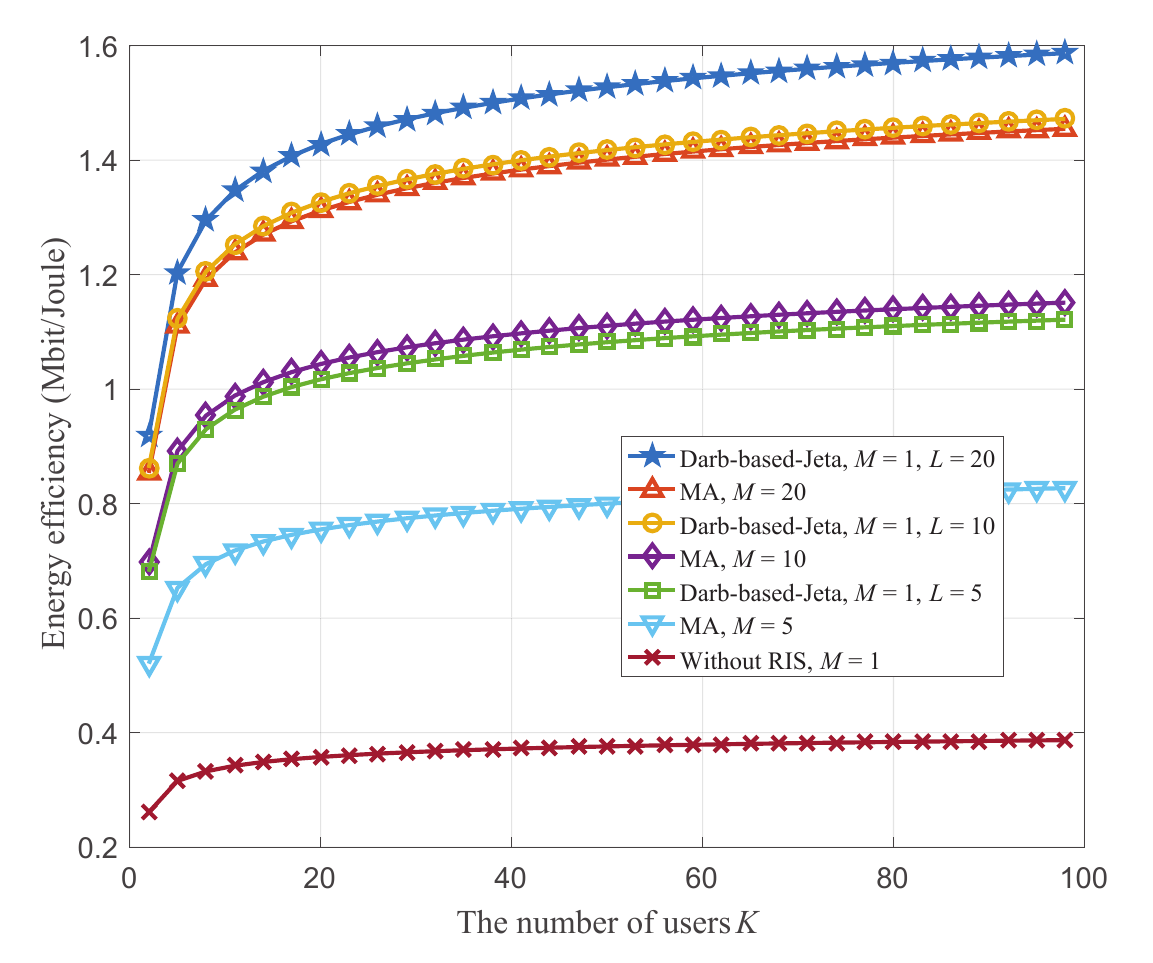}
\caption{The energy efficiency comparison of RIS-assisted and multi-antenna wireless communication system with different numbers of RIS elements and antennas.}\label{fig:F1}
\end{figure}

Figure~\ref{fig:F1} depicts the energy efficiency versus the number of users $K$ with different numbers of RIS elements and antennas. The energy efficiency increases as $K$ increases. This is because as the number of users $K$ in the system increases, the probability that each beam has a higher SINR among different users also increases which brings a multi-user diversity gain. For a fixed value of $K$, compared with the multi-antenna system without the help of RIS, the energy efficiency is larger by using the Darb-based-Jeta scheme. This means multiple RIS elements at the RIS-assisted transmitter can achieve the same function as multiple antennas and provide diversity gain for the system. 

\begin{figure}[htbp]
\centering\includegraphics[width=3.4in]{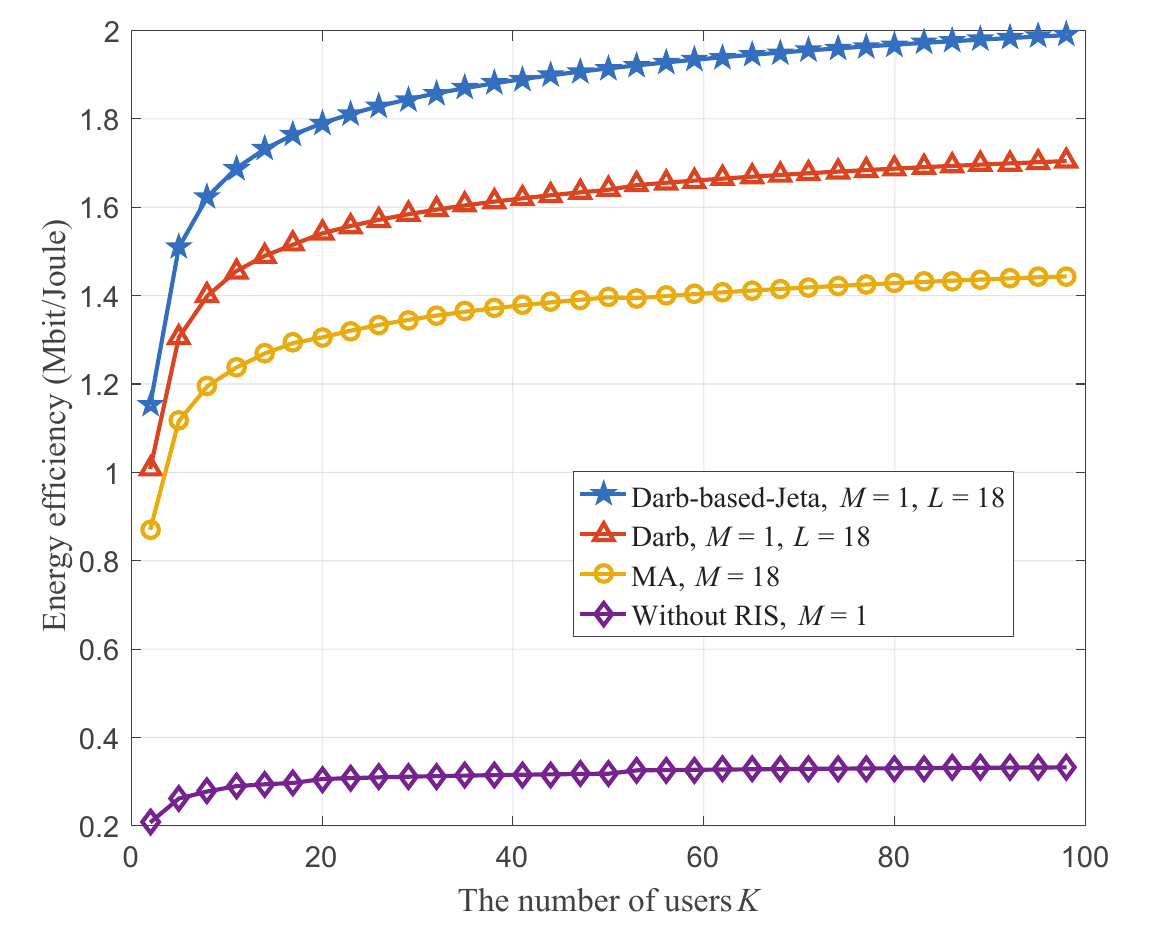}
\caption{The energy efficiency comparison of RIS-assisted with Darb scheme, Darb-based-Jeta scheme and multi-antenna wireless communication system.}\label{fig:F2}
\end{figure}

Figure~\ref{fig:F2} plots the energy efficiency versus the number of users $K$ in RIS-assisted and multi-antenna wireless communication system. Using our proposed Darb-based-Jeta scheme, the energy efficiency of the RIS-assisted system increases by jointly optimizing the number of RIS elements and the transmit power. As a result, the maximum energy efficiency is achieved when $L=18$ and $P_T=1.14$dBW. In addition, for a fixed value of $K$, the energy efficiency of the RIS-assisted system is larger than the multi-antenna system when we set $L=M$. This is due to the reason that RIS-assisted transmitter only needs single-active-antenna, which reduces the number of DACs, mixers and filters at transmitter. Therefore, the energy consumption of the system is well reduced and the energy efficiency correspondingly increases.

\begin{figure}[htbp]
\centering\includegraphics[width=3.6in]{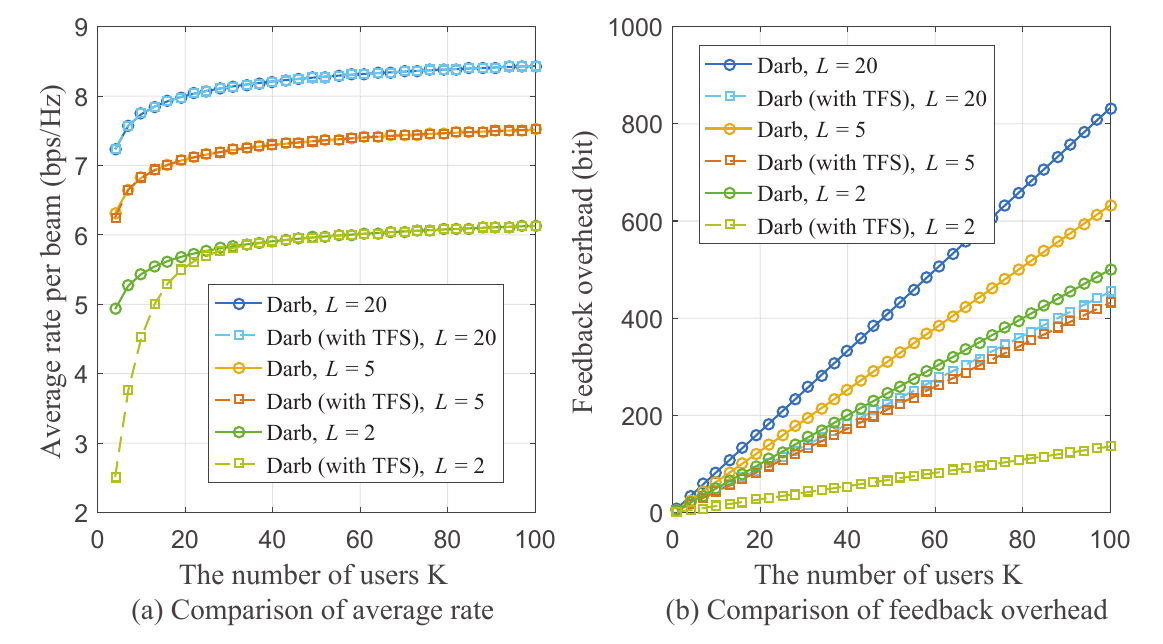}
\caption{The average rate and feedback overhead comparison of the selective-threshold feedback strategy.}\label{fig:F3}
\end{figure}

Figure~\ref{fig:F3} evaluates the average rate and feedback overhead versus the number of users $K$ under no threshold set and a threshold set with $\alpha = 0.1$. It can be seen from Fig.~\ref{fig:F3} (a) that when the numbers of users and RIS elements are both small, there is a gap between the average rate of our proposed scheme with the threshold feedback strategy (TFS) and without the TFS. However, when the number of users exceeds $20$, the average rate of the above two is almost the same, which means TFS does not affect the average rate. It can be seen from Fig.~\ref{fig:F3} (b) that feedback overhead is significantly reduced with the help of TFS, which means as long as the number of users in the system is large, the strategy can reduce feedback overhead without impacting the average rate.

\section{Conclusions}\label{sec:conc}
We solved the problems on how to maximize the energy efficiency of the RIS-assisted wireless communication system where the full CSI via RIS is very challenging to be obtained. First, we proposed a RIS-assisted transmitter structure with single-active-antenna. On this basis, we developed the Darb-based-Jeta scheme with a threshold feedback strategy, where RIS only needs to perform random phase shift without knowing the full CSI. We jointly optimized the number of RIS elements and the transmit power to enhance the energy efficiency of RIS-assisted wireless communications. Compared with the traditional multi-antenna system, our proposed scheme can significantly increase energy efficiency without full CSI.

\bibliographystyle{IEEEtran}
\bibliography{References}
\end{document}